\documentclass[a4paper]{article}

\usepackage{INTERSPEECH2020}
\usepackage{color}
\usepackage{multirow}
\usepackage{makecell}

\newcommand{\netname}{ContextNet}
\title{\netname: Improving Convolutional Neural Networks for Automatic Speech Recognition with Global Context}
\name{Wei Han\sthanks{~~Equal contribution.}, Zhengdong Zhang$^*$, Yu Zhang, Jiahui Yu, Chung-Cheng Chiu, James Qin, Anmol Gulati, Ruoming Pang, Yonghui Wu}

\address{Google Inc.}
\email{\{weihan, zhangzd\}@google.com}

\begin{document}

\maketitle
\begin{abstract}
Convolutional neural networks (CNN) have shown promising results for end-to-end speech recognition, albeit still behind RNN/transformer based models in performance. In this paper, we study how to bridge this gap and go beyond with a novel CNN-RNN-transducer architecture, which we call \netname. \netname~features a fully convolutional encoder that incorporates global context information into convolution layers by adding squeeze-and-excitation modules. In addition, we propose a simple scaling method that scales the widths of \netname{} that achieves good trade-off between computation and accuracy. 

We demonstrate that on the widely used Librispeech benchmark, \netname{} achieves a word error rate (WER) of 2.1\%/4.6\% without external language model (LM), 1.9\%/4.1\% with LM and 2.9\%/7.0\% with only 10M parameters on the clean/noisy LibriSpeech test sets. This compares to the best previously published model of 2.0\%/4.6\% with LM and 3.9\%/11.3\% with 20M parameters. The superiority of the proposed \netname{} model is
 also verified on a much larger internal dataset. 

\end{abstract}

\vspace{-0.1in}
\section{Introduction}

Convolution Neural Network (CNN) based models for end-to-end (E2E) speech recognition is attracting an increasing amount of attention~\cite{zhang2017very,zeghidour2018fully,li2019jasper,kriman2019quartznet}. Among them, the Jasper model~\cite{li2019jasper} recently achieves close to the state-of-the-art word error rate (WER) 2.95\%  on LibriSpeech test-clean~\cite{panayotov2015librispeech} with an external neural language model. The main feature of the Jasper model is a deep convolution based encoder with stacked layers of 1D convolutions and skip connections. Depthwise separable convolutions~\cite{chollet2017xception}  have been utilized to further increase the speed and accuracy of CNN models~\cite{hannun2019sequence,kriman2019quartznet}. The key advantage of a CNN based model is its parameter efficiency; however, the WER achieved by the best CNN model, QuartzNet~\cite{kriman2019quartznet}, is still behind the RNN/transformer based models~\cite{largespecaugment,karita2019comparative,wang2019transformer,zhang2020transformer}. 

A major difference between the RNN/Transformer~\cite{karita2019comparative, wang2019transformer, zhang2020transformer} based models and a CNN model is the length of the context. In a bidirectional RNN model, a cell in theory has access to the information of the whole sequence; in a Transformer model, the attention mechanism explicitly allows the nodes at two distant time stamps to attend each other. However, a naive convolution with a limited kernel size only covers a small window in the time domain; hence the context is small and the global information is not incorporated. In this paper, we argue that the lack of global context is the main cause of the gap of WER between the CNN based ASR model and the RNN/Transformer based models. 

To enhance the global context in the CNN model, we draw inspirations from the squeeze-and-excitation (SE) layer introduced in~\cite{hu2018squeeze}, and propose a novel CNN model for ASR, which  we call \netname. An SE layer squeezes a sequence of local feature vectors into a single global context vector, broadcasts this context back to each local feature vector, and merges the two via multiplications. When we place an SE layer after a naive convolution layer, we grant the convolution output the access to global information. Empirically, we observe that adding squeeze-and-excitation layers to \netname~introduces the most reduction in the WER on LibriSpeech test-other.

Previous works on hybrid ASR have successfully introduced the context to acoustic models by either stacking a large number of layers, or having a separately trained global vector to represent the speaker and the environment information \cite{peddinti2015jhu,xue2014fast,karafiat2011ivector,saon2013speaker}. In \cite{sailor2019unsupervised}, SE has been adopted to RNN for unsupervised adaptation. In this paper, we show that SE can also be effective for CNN encoders.

The architecture of \netname{} is also inspired by the design choices of QuartzNet~\cite{kriman2019quartznet}, such as the usage of depthwise separable 1D convolution in the encoder. However, there are some key differences in the architectures in addition to the incorporation of the SE layer. For instance, we use a RNN-T decoder~\cite{graves2012sequence, rao2017exploring, he2019, tara2020} instead of the CTC decoder~\cite{graves2006connectionist}. Moreover, we use the Swish activation function~\cite{ramachandran2017searching}, which contributes a slight but consistent reduction in WER. Overall, \netname{} achieves the WER of 1.9\%/4.1\% on LibriSpeech test-clean/test-other. This is a big improvement over previous CNN based architectures such as QuartzNet\cite{kriman2019quartznet}, and it outperforms transformer and LSTM based models~\cite{zhang2020transformer, wang2019transformer, zeyer2019comparison, karita2019comparative, park2019specaugment}.

This paper also studies how to reduce the computation cost of \netname{} for faster training and inference. First, we adopt a progressive downsampling scheme that is commonly used in vision models. Specifically, we progressively reduce the length of the encoded sequence eight times, significantly lower the computation while maintaining the encoder's representation power and the overall model accuracy. As a benefit, this downsampling scheme allows us to reduce the kernel size of all the convolution layers to five without significantly reducing the effective receptive field of an encoder output node.

\begin{figure}[t]
\begin{centering}
\includegraphics[width=1\columnwidth]{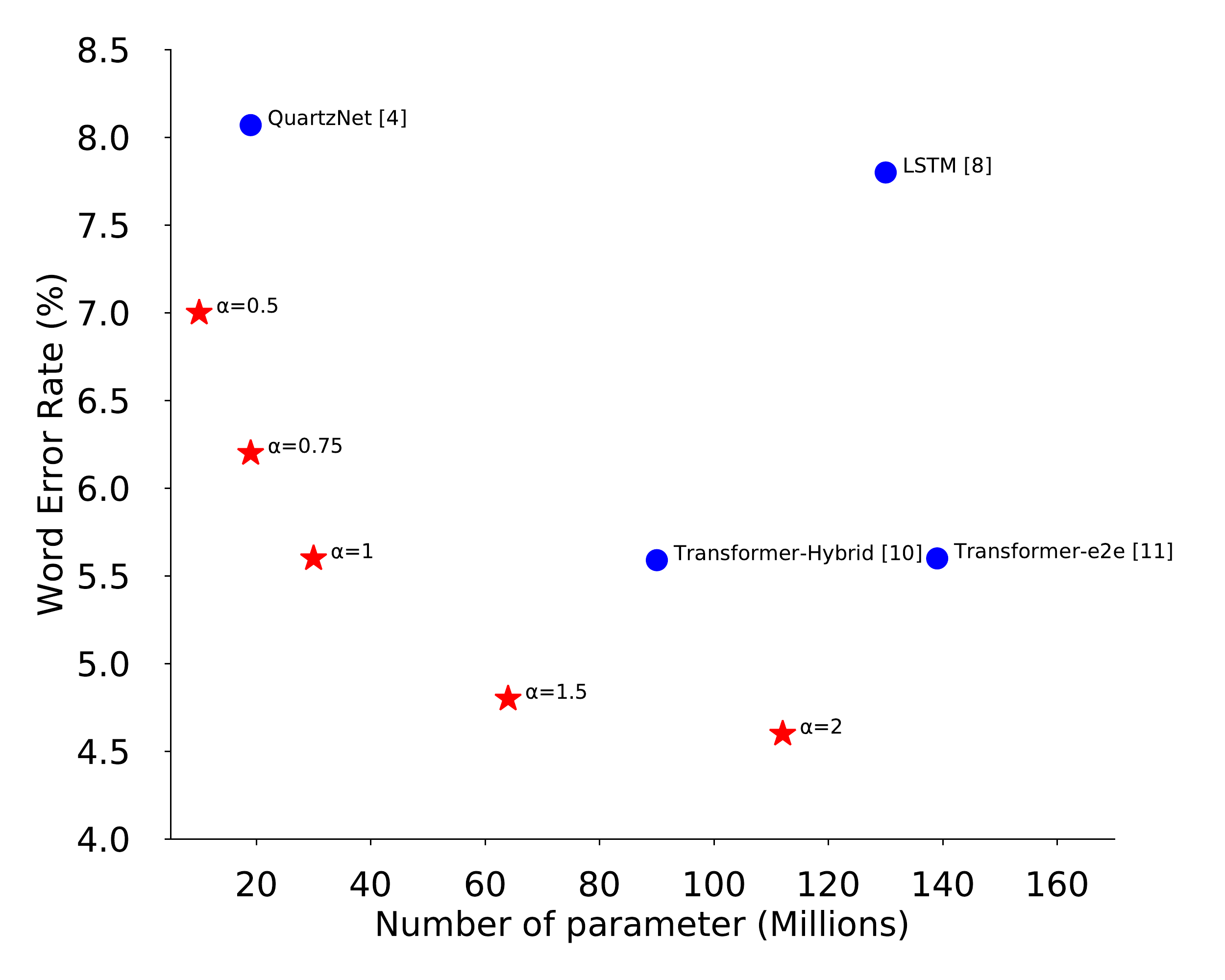}
\par\end{centering}
\caption{LibriSpeech test-other WER vs.\ model size. All numbers for E2E models are without external LM. For the Transformer-Hybrid model we use the encoder size. \netname{} numbers are highlighted in red, and $\alpha$ is the model scaling parameter discussed in Section~\ref{sec:model:details}. Detailed results are in Table \ref{tab:libri_speech_main}.}
\label{overall}
\end{figure}

We can scale \netname{} by globally changing the number of channels in convolutional filters. Figure~\ref{overall} illustrates the trade-off of {\netname} between model size and WER, as well as its comparison against other methods. Clearly, our scaled model achieves the best trade-offs among all.

In summary, the main contributions of this paper are: (1) an improved CNN architecture with global context for ASR, (2) a progressive downsampling and model scaling scheme to achieve superior accuracy and model size trade-off.
\vspace{-0.05in}
\section{Model}
\vspace{-0.05in}
This section introduces the architecture details of \netname. Section~\ref{sec:model:transducer} discusses the high-level design of \netname. Then Section~\ref{sec:model:encoder} introduces our convolutional encoder, and discusses how we progressively reduce the temporal length of the input utterance in the network to reduce the computation while maintaining the accuracy of the model.  
\vspace{-0.05in}
\subsection{End-to-end Network: CNN-RNN-Transducer}
\label{sec:model:transducer}
\vspace{-0.02in}
Our network is based on the RNN-Transducer framework \cite{graves2012sequence, rao2017exploring, he2019}. The network contains three components: audio encoder on the input utterance, label encoder on the input label, and a joint network to combine the two and decode. We directly use the LSTM based label encoder and the joint network from \cite{he2019}, but propose a new CNN based audio encoder. 
\vspace{-0.05in}
\subsection{Encoder Design}
\vspace{-0.05in}
\label{sec:model:encoder}
Let the input sequence be $\mathbf{x}=(x_1, \ldots, x_T)$. The encoder transforms the original signal $\mathbf{x}$ into a high level representation $\mathbf{h}=(h_1, \ldots, h_{T'})$, where $T'\leq T$. Our convolution based $\mathrm{AudioEncoder}(\cdot)$ is defined as:
\begin{equation}
\begin{small}
    \mathbf{h} = \mathrm{AudioEncoder}(\mathbf{x}) = \mathrm{C}_K\left(\mathrm{C}_{K-1}\left(\ldots \mathrm{C}_{1}(\mathbf{x})\right)\right),
\label{eqn:encoder}
\end{small}
\end{equation}
where each $\mathrm{C}_k(\cdot)$ defines a convolution block. It contains a few layers of convolutions, each followed by batch normalization~\cite{goodfellow2016deep} and an activation function. It also includes the squeeze-and-excitation component~\cite{hu2018squeeze} and skip connections~\cite{he2016deep}. 

Before presenting the details of $\mathrm{C}(\cdot)$, we first elaborate the important modules in $\mathrm{C}(\cdot)$.

\vspace{-0.05in}
\subsubsection{Squeeze-and-excitation}
\begin{figure*}
    \centering
    \includegraphics[width=0.8\textwidth]{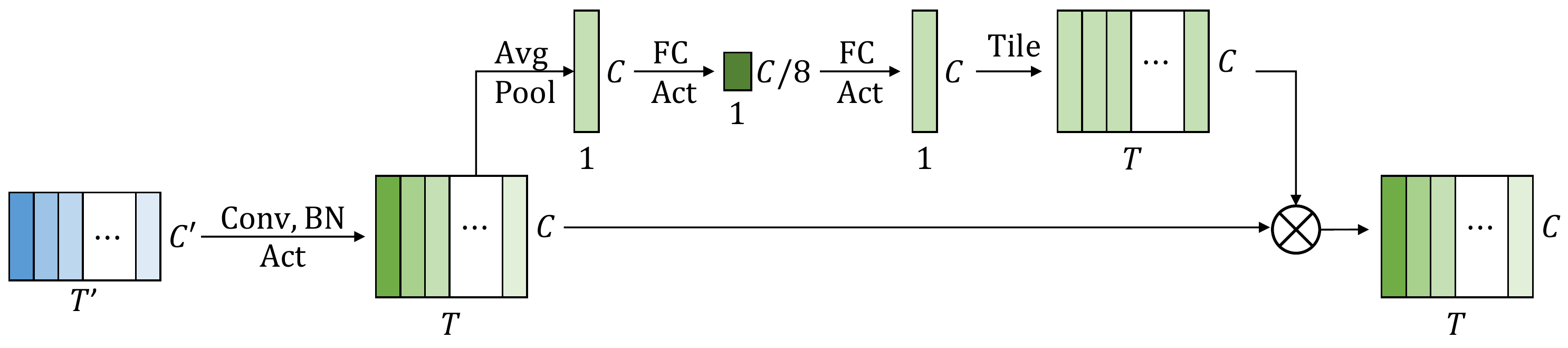}
    \caption{1D Squeeze-and-excitation module. The input first goes through a convolution layer followed by batch normalization and activation. Then average pooling is applied to condense the conv result into a 1D vector, which is then processed by a bottleneck structure formed by two fully connected (FC) layers with activation functions. The output goes through a Sigmoid function to be mapped to $(0, 1)$, and then tiled and applied on the conv output using pointwise multiplications.}
    \label{fig:squeeze_and_excitation}
\end{figure*}

\vspace{-0.05in}
As illustrated in Figure~\ref{fig:squeeze_and_excitation}, the Squeeze-and-excitation~\cite{hu2018squeeze} function, $\mathrm{SE}(\cdot)$, performs global average pooling on the input $x$, transforms it into a global channelwise weight $\theta(x)$ and element-wise multiplies each frame by this weight. We adopt the idea to the 1D case,
\begin{small}
\begin{equation*}
\begin{split}
    \bar{x}=\frac{1}{T} \sum_{t} x_t,&~\theta(x)=\mathrm{Sigmoid}(W_2(\mathrm{Act}(W_1\bar{x} + b_1)) + b_2), \\
    &\mathrm{SE}(x)=\theta(x)\circ x, 
\end{split}
\end{equation*}
\end{small}
where $\circ$ represents element-wise multiplication, $W_1, W_2$ are weight matrics, and $b_1, b_2$ are bias vectors. 

\vspace{-0.05in}
\subsubsection{Depthwise separable convolution}
\vspace{-0.02in}
Let $\mathrm{conv}(\cdot)$ represent the convolution function used in the encoder.
In this paper, we choose depthwise separable convolution as $\mathrm{conv}(\cdot)$, because such a design has been previously shown in various applications \cite{chollet2017xception,kriman2019quartznet,sandler2018mobilenetv2} to achieve better parameter efficiency without impacting accuracy.

For simplicity, we use the same kernel size on all depthwise convolution layers in the network.

\vspace{-0.05in}
\subsubsection{Swish activation function}
\vspace{-0.05in}
Let $\mathrm{Act}(\cdot)$ represent the activation function in the encoder. To choose $\mathrm{Act}(\cdot)$, we've experimented with both ReLU and the swish function~\cite{ramachandran2017searching} defined as:
\begin{small}
\begin{equation}
\mathrm{Act}(x)=x\cdot \sigma(\beta x)=\frac{x}{1+\exp{\left(-\beta x\right)}},
\label{eq:swish}
\end{equation}
\end{small}
where $\beta=1$ for all our experiments. We've observed that the swish function works consistently better than ReLU.  
\vspace{-0.05in}
\subsubsection{Convolution block}
\begin{figure}
    \centering
    \includegraphics[width=0.85\columnwidth]{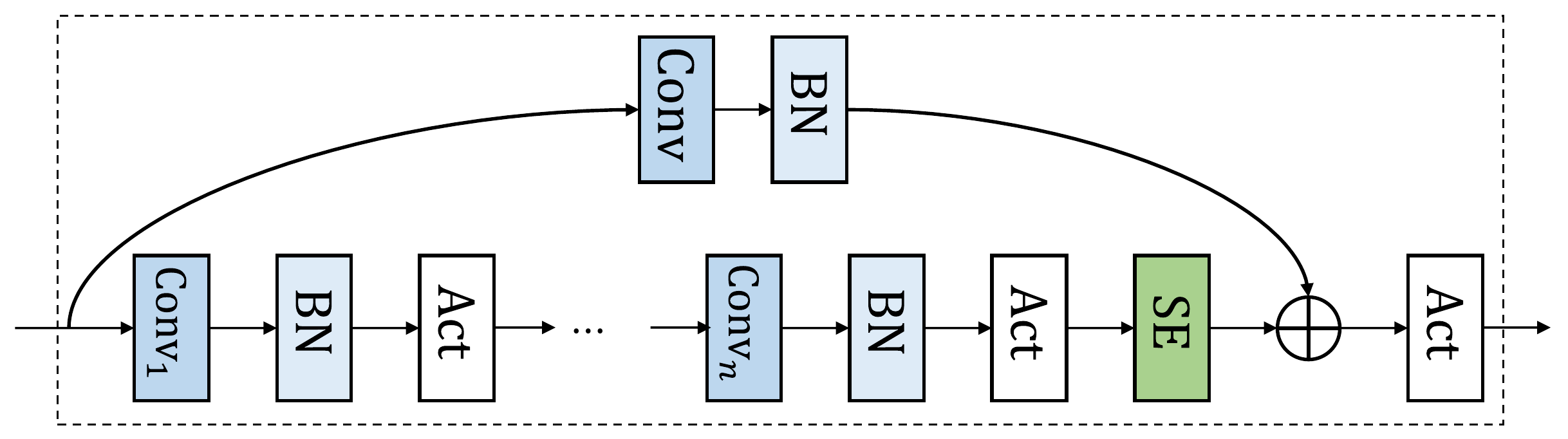}
    \caption{A convolution block $C_i$ contains a number of convolutions, each followed by batch normalization and activation. A squeeze-and-excitation (SE) block operates on the output of the last convolution layer. A skip connection with projection is applied on the output of the squeeze-and-excitation block. }
    \label{fig:conv_block}
\end{figure}

\vspace{-0.02in}
With all the individual modules introduced, we now present the convolution block $C(\cdot)$ from Equation \eqref{eqn:encoder}. Figure~\ref{fig:conv_block} illustrate a high-level architecture of $C(\cdot)$. A block $C(\cdot)$ can contain a few $\mathrm{Conv}(\cdot)$ functions; let $m$ be the number of $\mathrm{Conv}(\cdot)$ functions. Let $BN(\cdot)$ be the batch normalization~\cite{ioffe2015batch}. We define each layer as $f(x)=\mathrm{Act}(\mathrm{BN}(\mathrm{Conv}(x))$. Therefore, 

\begin{small}
\begin{equation*}
\mathrm{C}(x) = \mathrm{Act}\left(\mathrm{SE}\left(f^m(x)\right) + \mathrm{P}(x)\right)
\end{equation*}
\end{small} where $f^m$ means stacking $m$ layers of the function $f(\cdot)$ on the input and $\mathrm{P}(\cdot)$ represents a pointwise projection function on the residual. By a slight abuse of notation, we allow the first layer and the last layer to be different from the other $m-2$ layers: if the block needs to downsample the input sequence by two times, the last layer has a stride of two while all the rest $m-1$ layers has a stride of one; otherwise all $m$ layers have a stride of one. Additionally, if the block has an input number of channels $D_{in}$ and output number of channels $D_{out}$, the first layer $f'$ turns $D_{in}$ channels into $D_{out}$ channels while the rest $m-1$ layers maintain the number of channels as $D_{out}$. Following the convention, the projection function $\mathrm{P}$ has the same number of stride as the first layer.

\vspace{-0.05in}
\subsubsection{Progressive downsampling}
We use strided convolution for temporal downsampling. More downsampling layers reduces computation cost, but excessive downsampling in the encoder may negatively impact the decoder. Empirically, we find that a progressive $8\times$ downsampling scheme achieves a good trade-off between speed and accuracy. These trade-offs are discussed in Section~\ref{sec:exp:filter_size}.

\vspace{-0.05in}
\subsubsection{Configuration details of \netname}
\vspace{-0.02in}
\label{sec:model:details}
\netname{} has $23$ convolution blocks $C_0, \ldots, C_{22}$. All convolution blocks have five layers of convolution, except $C_0$ and $C_{22}$, which only have one layer of convolution each. Table \ref{tab:network_configuration} summarizes the architecture details.
Note that a global parameter $\alpha$ controls the scaling of our model. Increasing $\alpha$ when $\alpha > 1$ increases the number of channels of the convolutions, giving the model more representation power with a larger model size.

\begin{table}[!tb]
    \centering
    \caption{Configuration of the \netname{} encoder. $\alpha$ controls the number of output channels, and thus the scaling of our model. The kernel size is for the window size in the temporal domain; the convolutions are across frequency. If the stride of a convolution block is 2, its last conv layer has a stride of two while the rest of the conv layers has a stride of one, as discussed in Section~\ref{sec:model:encoder}.}
    \resizebox{0.9\columnwidth}{!}{
    \begin{tabular}{ccccc}
        \toprule
         \textbf{Block ID} & \bf \makecell{\#Conv \\ layers} & \bf \makecell{\#Output \\channels} & \bf \makecell{Kernel \\ size} & \bf Other \\ \midrule
         $C_0$ & 1 & $256\times\alpha$ & 5 & No residual \\ \midrule
         $C_1$-$C_2$ & 5 & $256\times\alpha$ & 5 & \\ 
         $C_3$ & 5 & $256\times\alpha$ & 5 & stride is $2$\\ \midrule
         $C_4$-$C_6$ & 5 & $256\times\alpha$ & 5 &\\ 
         $C_7$ & 5 & $256\times\alpha$ & 5 & stride is $2$\\ \midrule
         $C_8$-$C_{10}$ & 5 & $256\times\alpha$ & 5 &\\ 
         $C_{11}$-$C_{13}$ & 5 & $512\times\alpha$ & 5 & \\ 
         $C_{14}$ & 5 & $512\times\alpha$ & 5 & stride is $2$ \\ \midrule
         $C_{15}$-$C_{21}$ & 5 & $512\times\alpha$ & 5 & \\ 
         $C_{22}$ & 1 & $640\times\alpha$ & 5 & No residual\\
         \bottomrule
    \end{tabular}
    }
    \label{tab:network_configuration}
\end{table}

\vspace{-0.05in}
\section{Experiments}  
\vspace{-0.02in}
\label{sec:exp}

We conduct experiments on the Librispeech \cite{panayotov2015librispeech} dataset which consists of 970 hours of labeled speech and an additional text only corpus for building language model. We extract 80 dimensional filterbanks features using a 25ms window with a stride of 10ms.

We use the Adam optimizer~\cite{kingma2014adam} and a transformer learning rate schedule~\cite{vaswani2017attention} with 15k warm-up steps and a peak learning rate of $0.0025$. An $\ell_2$ regularization with $10^{-6}$ weight is also added to all the trainable weights in the network. We use a single layer LSTM as decoder with input dimension of 640.  Variational noise is introduced to the decoder as a regularization. 

We use SpecAugment~\cite{park2019specaugment, largespecaugment} with mask parameter ($F = 27$), and ten time masks with maximum time-mask ratio ($p_S = 0.05$), where the maximum size of the time mask is set to $p_S$ times the length of the utterance. Time warping is not used. We use a 3-layer LSTM LM with width 4096 trained on the LibriSpeech langauge model corpus with the LibriSpeech960h transcripts added, tokenized with the 1k WPM built from LibriSpeech 960h. The LM has word-level perplexity 63.9 on the dev-set transcripts. The LM weight $\lambda$ for shallow fusion is tuned on the dev-set via grid search. All models are implemented with Lingvo toolkit~\cite{lingvo}.

\vspace{-0.05in}
\subsection{Results on LibriSpeech}
\vspace{-0.02in}

We evaluate three different configurations of \netname{} on LibriSpeech. The models are all based  on Table~\ref{tab:network_configuration}, but differ in the network width, $\alpha$; hence, they differ in model size. Specifically, we choose $\alpha$ in $\{0.5, 1, 2\}$ for the small, medium and large \netname{}. We also build our own LSTM baseline as a reference.

Table \ref{tab:libri_speech_main} summarizes the evaluation results as well as the comparisons with a few previously published systems. The results suggest improvements of \netname{} over previously published systems. Our medium model, \netname(M), only has $31M$ parameters and achieves similar WER compared with much larger systems~\cite{zhang2020transformer,synnaeve2019endtoend}. The large model, \netname(L), outperforms the previous SOTA by 13\% relatively on test-clean and 18\% relatively on test-other. Our scaled-down model, \netname(S), also shows an improvement to previous systems of similar size~\cite{kriman2019quartznet}, with or without a language model. 

\begin{table}[!h]
  \caption{WER on Librispeech. Compared to previous models, {\netname}  achieves superior performance both with and without language models}
  \label{tab:libri_speech_main}
  \centering
  \resizebox{\columnwidth}{!}{%
  \begin{tabular}{lccccc}
    \toprule
    \bfseries Method & \#{\bfseries Params (M)} & \multicolumn{2}{c}{\bfseries Without LM} & \multicolumn{2}{c}{\bfseries With LM} \\
    \cmidrule(r){3-4} \cmidrule(r){5-6}
     & & \bfseries testclean & \bfseries testother & \bfseries testclean & \bfseries testother \\
    \midrule
    \bfseries Hybrid \\
    \quad Transformer~\cite{wang2019transformer} & - 
    & - & - & 2.26 & 4.85 \\
    \bfseries CTC \\
    \quad QuartzNet (CNN)~\cite{kriman2019quartznet} & 19
    &3.90 & 11.28 & 2.69 & 7.25\\
    \bfseries LAS \\
    \quad Transformer \cite{karita2019comparative} & -
    & - & - & 2.6 & 5.7 \\
    \quad  Transformer \cite{synnaeve2019endtoend} & 270
    & 2.89 & 6.98 & 2.33 & 5.17 \\
    \quad LSTM & 360
    & 2.6 & 6.0 & 2.2 & 5.2 \\
    \bfseries Transducer \\
    \quad Transformer \cite{zhang2020transformer} & 139
    & 2.4 & 5.6 & 2.0 & 4.6 \\
    \midrule
    \bfseries This Work\\
    \quad\netname(S) & 10.8
    & \textbf{2.9} & \textbf{7.0} & \textbf{2.3} & \textbf{5.5}\\
    \quad\netname(M) & 31.4
    & \textbf{2.4} & \textbf{5.4} & \textbf{2.0} & \textbf{4.5}\\
    \quad\netname(L) & 112.7
    & \textbf{2.1} & \textbf{4.6} & \textbf{1.9} & \textbf{4.1} \\
    \bottomrule
  \end{tabular}
  }
  \vskip -0.1in
\end{table}

\vspace{-0.05in}
\subsection{Effect of Context Size}
\vspace{-0.02in}
To validate the effectiveness of adding global context to the CNN model for ASR, we perform an ablation study on how the squeeze-and-excitation module affects the WER on LibriSpeech test-clean/test-other. \netname~in Table~\ref{tab:network_configuration} with all squeeze-and-excitation modules removed and $\alpha=1.25$ serves as the baseline of zero context.

The vanilla squeeze-and-excitation module uses the whole utterance as context. To investigate the effect of different context sizes, we replace the global average pooling operator of the squeeze-and-excitation module by a stride-one pooling operator where the context can be controlled by the size of the pooling window. In this study, we compare the window size of $256$, $512$ and $1024$ on all convolutional blocks.

As illustrated in Table~\ref{tab:context_size}, the SE module provides major improvement over the baseline. In addition, the benefit becomes greater as the length of the context window increases. This is consistent with the observation in a similar study of SE on image classification models~\cite{hu2018gather}.

\begin{table}[h!]
\centering
\caption{Effect of the context window size on WER. All models have $\alpha=1.25$.}

\resizebox{0.7\columnwidth}{!}{
\begin{tabular}{ccccc}
\toprule 
\bfseries Context & \bf \makecell{dev\\ clean} & \bf \makecell{dev\\ other} & \bf \makecell{test\\ clean} & \bf \makecell{test\\ other}\\
\midrule 
None & 2.6 & 7,0 & 2.6 & 6.9 \tabularnewline
256 & 2.1 & 5.4 & 2.3 & 5.5\tabularnewline
512 & 2.1 & 5.1 & 2.3 & 5.2\tabularnewline
1024 & 2.1 & 5.0 & 2.3 & 5.1\tabularnewline
global & \textbf{2.0} & \textbf{4.9} & \textbf{2.3} & \textbf{4.9}\\
\bottomrule
\end{tabular}
}
\label{tab:context_size}
\end{table}

\vspace{-0.1in}
\subsection{Depth, Width, Kernel Size and Downsampling}
\vspace{-0.02in}
\label{sec:exp:filter_size}
\textit{Depth:\ } We perform a sweeping on the number of convolutional blocks and our best configuration is in Table~\ref{tab:network_configuration}. We find that with this configuration, we can train a model in a day with stable convergence.

\begin{table}
\centering
\caption{The effect of temporal reduction and convolution kernel size on FLOPS and model accuracy.}
\label{tab:ablation_filter}
\resizebox{0.9\columnwidth}{!}{
\begin{tabular}{ccccccc}
\toprule 
\bf Reduction & \makecell{\bf Kernel\\ \bf size} & \bf GFLOPS\footnotemark  & \bf testclean & \bf testother \\
\midrule 
2x & 3 & 2.131 &  2.7 & 6.3 & \tabularnewline
 & 5 & 2.137  &  2.6 & 5.8 & \tabularnewline
 & 11 & 2.156 & 2.4 & 5.4 & \tabularnewline
 & 23 & 2.194 & 2.3 & 5.0 & \tabularnewline
 \midrule
8x & 3 & 1.036 & 2.3 & 5.1 \tabularnewline
 & 5 & 1.040 &  2.3 & 5.0 \tabularnewline
 & 11  &  1.050 & 2.3 & 5.0 \tabularnewline
 & 23  &  1.071 & 2.3 & 5.2 \tabularnewline
\bottomrule 
\end{tabular}
}
\par
\end{table}

\begin{table}[h!]
\centering
\caption{Effect of Model scaling by network width on WER.}
\resizebox{0.85\columnwidth}{!}{
\begin{tabular}{cccccc}
\toprule
$\mathbf{\alpha}$ & \bf \#Params(M) & \bf \makecell{dev\\ clean} & \bf \makecell{dev\\ other} & \bf \makecell{test\\ clean} & \bf \makecell{test\\ other} \\
\midrule 
0.5 & 10.8 & 2.7 & 7.0 & 2.9 & 7.0 \tabularnewline
1 & 31.4 & 2.2 & 5.1 & 2.4 & 5.4 \tabularnewline
1.5 & 65.4 & 2.0 & 4.7 & 2.2 & 4.8 \tabularnewline
2 & 112.7 & 2.0 & 4.6 & 2.1 & 4.6\tabularnewline
\bottomrule 
\end{tabular}
}
\label{tab:ablation_width}
\end{table}

\vspace{0.1in}
\noindent \textit{Width:\ \ }\label{sec:term:width} We globally scale the width of the network (i.e., the number of channels) on all encoder layers and study how it impacts the model performance. Specifically, we take the \netname{} model from Table~\ref{tab:network_configuration}, sweep $\alpha$, and report the model size and the WER on LibriSpeech. Table \ref{tab:ablation_width} summarizes the result; it demonstrates the good trade-off between model size and WER of \netname{}.

\footnotetext{We report the average encoder FLOPS for processing one second of audio.}

\vspace{0.1in}
\noindent
\textit{Downsampling and kernel size:\ }
Table \ref{tab:ablation_filter} summarizes the FLOPS and WER on LibriSpeech with various choices of downsampling and fileter size. We use the same model with only one downsampling layer added to $C_3$ as the baseline; hence the baseline only does $2\times$ temporal reduction. We sweep the kernel size in $\{3, 5, 11, 21\}$, each kernel size is applied to all the depthwise convolution layers. The results suggest that progressive downsampling introduces significant saving in the number of FLOPS. Moreover, it actually benefits the accuracy of the model slightly. In addition, with progressive downsampling, increasing the kernel size decreases the WER of the model.

\vspace{-0.05in}
\subsection{Large Scale Experiments}
\vspace{-0.02in}
Finally, we show that the proposed architecture is also effective on large scale datasets. We use a experiment setup similar to \cite{chiu20longform}, where the training set has public Youtube videos with semi-supervised transcripts generated by the approach in \cite{liao2013large}. We evaluate on 117 videos with a total duration of 24.12 hours. This test set has diverse and challenging acoustic environments \footnote{Reproduced results. The train and eval set has been changed recently so the numbers in Table \ref{tab:youtube} are different from reported in \cite{chiu20longform}.}. Table~\ref{tab:youtube} summarizes the result. We can see that \netname~ outperforms the previous best architecture from \cite{chiu20longform}, which is a combination of convolution and bidirectional LSTM, by 12\% relatively with fewer parameters and FLOPS.

\begin{table}[h!]
\centering
\caption{Comparing {\netname} with previous best results on Youtube test sets}
\resizebox{0.85\columnwidth}{!}{
\begin{tabular}{cccc}
\toprule 
\bf Model & \bf \#Params (M) & \bf GFLOPS  & \bf Youtube WER \tabularnewline
\midrule 
TDNN \cite{chiu20longform} & 192 & 3.834 & 9.3 \tabularnewline
\netname & 112 & 2.647 & 8.2 \tabularnewline
\bottomrule 
\end{tabular}}
\label{tab:youtube}
\end{table}

\vspace{-0.2in}
\section{Conclusion}
\vspace{-0.05in}

In this work, we proposed and evaluated a CNN based architecture for end-to-end speech recognition. A couple of modeling choices are discussed and compared. This model achieves a better accuracy on the LibriSpeech benchmark with much fewer parameters compared to previously published CNN models. The proposed architecture can easily be used to search for small ASR models by limiting the width of the network. Initial study on a much larger and more challenging dataset also confirms our findings.

\bibliographystyle{IEEEtran}

\bibliography{mybib}

\end{document}